\documentclass[aps,reprint]{revtex4-2}
\usepackage{epsfig}% 
\usepackage{epstopdf}
\usepackage{multirow}
\usepackage{graphicx}% 
\usepackage{dcolumn}% 
\usepackage{bm}% bold math
\usepackage{textcomp}
\usepackage{array}
\usepackage{csquotes}
\usepackage{subcaption}
\usepackage{booktabs}
\usepackage{amsmath}
\usepackage{hyperref}
\hypersetup{
    colorlinks=true,
    linkcolor=blue,
    filecolor=magenta,      
    urlcolor=red,
    citecolor=red,
    pdftitle={},
    }
    
\begin{document}
\title{Strain-tunable type-II to type-III \& Gimbal nodal line transition in\\ $Imm2$-phase of Cu$_2$SnS$_3$: An \textit{ab-initio} study}
\author{Prakash Pandey$^{1}$}
\altaffiliation{ \url{prakashpandey6215@gmail.com}}
\author{Sudhir K. Pandey$^{2}$}
\altaffiliation{ \url{sudhir@iitmandi.ac.in}}
\affiliation{$^{1}$School of Physical Sciences, Indian Institute of Technology Mandi, Kamand - 175075, India\\$^{2}$School of Mechanical and Materials Engineering, Indian Institute of Technology Mandi, Kamand - 175075, India}

\date{\today}

\begin{abstract}
Topological nodal line semimetals (NLSMs) represent an intriguing quantum phase, opening new avenues in materials science for practical applications such as anisotropic transport devices, high-mobility conductors, unconventional thermoelectrics, and nonlinear optical devices. Recently, Cu$_2$SnS$_3$ has been theoretically proposed as a type-II NLSM, with its Fermi surface containing only one nodal ring. Here, we demonstrate how uniaxial, equi-biaxial, and equi-triaxial strains affect the nodal line state of the $Imm2$-phase of Cu$_2$SnS$_3$ by using \textit{state-of-the-art ab-initio} calculations. Under the application of uniaxial compressive strain (UCS) along the \textit{a}-direction, the plane of the nodal line evolves from the $k_x$-$k_z$ to $k_y$-$k_z$ for 6\%$\leq$UCS$\leq$8\%. In contrast, under uniaxial tensile strain (UTS), the nodal line remains in the ($k_x$-$k_z$) plane across the entire studied range of UTS. Interestingly, on the application of equi-biaxial tensile strain (EBTS) along \textit{a}-\textit{b} (\textit{a-c}) directions, it hosts only one nodal ring below 8\% ($<$6\%), which further evolves into three (seven) nodal-ring for EBTS of 8\% (6\%$\leq$EBTS$\leq$8\%). 
Upon the application of EBTS along \textit{a-c} directions, we found two sets of three mutually orthogonal, intersecting nodal loops (topological gimbals). 
Apart from this, under the application of equi-biaxial compressive strain (EBCS) along the \textit{a-b} (\textit{a-c}) directions, it exhibits only one nodal ring up to 8\% (7\%). Beyond this, the nodal line completely vanishes and does not reappear at higher values of EBCS.
Under equi-triaxial tensile strain (ETTS), Cu$_2$SnS$_3$ exhibits only one nodal-ring $<$6\%, which subsequently transforms into five nodal-ring for 6\%$\leq$ETTS$\leq$8\%. However, under the application of equi-triaxial compressive strain (ETCS), as in EBCS, only one nodal line exists up to 6\% ETCS. After this threshold, it vanishes completely and does not emerge again at higher ETCS values. 
This study unveils that external strain can modulate the topological properties of the nodal line, making it a versatile platform for applications such as anisotropic transport, gated topological transistors, quantum gyroscopes, and anomalous magnetotransport.

\end{abstract}

\maketitle

%******************************************************** Introduction ************************************************************
%\setlength{\parindent}{3em}
\section{Introduction}
%{\it Introduction.|}
%%%%%Basic-details about topological semimetals
Topological materials have recently been identified as new phases of quantum matter that exist outside the scope of Landau’s theory of phase transitions~\cite{RevModPhys.89.040502, RevModPhys.88.035005}. These materials are of both scientific and technological interest, and they are classified into topological semimetals \& topological insulators~\cite{10.1126/science.1133734, PhysRevB.85.195320, PhysRevB.85.195320, PhysRevB.88.125427, PhysRevX.5.011029, PhysRevX.5.031013}. 
Nodal line semimetals (NLSMs) are a type of topological semimetal in which the highest occupied and lowest unoccupied bands form a closed path in 3D momentum space near the Fermi level (E$_{\rm{F}}$)~\cite{RevModPhys.90.015001, Fang_2016}.
Such loops are often known as nodal lines, nodal rings~\cite{PhysRevB.93.205132, PhysRevB.92.081201, PhysRevB.84.235126}, nodal chains~\cite{PhysRevLett.119.036401, yan2018experimental, bzduvsek2016nodal}, nodal links~\cite{PhysRevB.96.041102, PhysRevB.96.041103}, and nodal knot~\cite{PhysRevB.96.201305}. 
%%%%%Application point of View
Among topological semimetals, NLSMs exhibit unique topological features that can give rise to several intriguing phenomena, including robust topological surface states, a nontrivial Berry phase, and the spin Hall effect~\cite{PhysRevB.90.115111, PhysRevB.84.235126, PhysRevB.92.081201, 10.1126/sciadv.1603266, PhysRevB.83.220503, PhysRevB.95.235104}.
In addition to this, NLSMs show nearly flat drumhead-like surface states, paving the way for high-temperature superconductivity~\cite{PhysRevB.83.220503, Volovik_2015} and presenting an intriguing prospect for correlation physics~\cite{PhysRevB.93.035138, PhysRevB.95.075426}. Furthermore, the appearance of exact zero modes due to almost nondispersive Landau levels at E$_{\rm{F}}$ has been reported to exist inside the nodal ring~\cite{PhysRevB.92.045126}. The NLSMs exhibit a quasitopological electromagnetic response that is directly linked to charge polarization and orbital magnetization~\cite{PhysRevB.95.075138}. 
Other than the aforementioned novel physical properties, they also offer special collective mode~\cite{PhysRevB.93.085138}, anisotropic electronic transport~\cite{PhysRevLett.115.026403}, and unusual optical responses~\cite{Carbotte_2017}.
Despite the importance of the nodal line from a physics point of view, it has potential applications for several fields in solid-state physics and material science. For example, NLSMs have been predicted to be potential catalysts in electrocatalytic processes for hydrogen evolution reactions~\cite{li2024high}. 
In addition to this, the topological NLSMs have been proposed as a candidate for efficient thermoelectric converters~\cite{PhysRevB.105.085406}. Other applications of NLSMs are used in ultra-high-speed devices (due to their low energy consumption and quantum transport), phototransistors, and manufacturing sonic devices~\cite{zou2023topological, yang2018symmetry, PhysRevB.106.184101}.

%%% How to maintain nodal line against splitting

Despite numerous theoretical and experimental studies, only a few NLSMs that show only one nodal ring have undergone experimental verification~\cite{PhysRevB.93.205132, 10.1063/1.4926545, PhysRevB.92.045108, PhysRevLett.116.195501, chen2015nanostructured, PhysRevB.94.195104, PhysRevB.95.045136, PhysRevLett.118.176402, PhysRevLett.131.236903, yamakage2016line, PhysRevB.96.115106}. In literature, it is reported that many materials show NLSMs in the absence of spin-orbit coupling (SOC)~\cite{Pandey_2023, PhysRevResearch.4.033067, Fang_2016, PhysRevLett.119.036401}. 
However, after considering the effect of SOC, nodal line may either gapped into several nodal points, influenced by the crystalline symmetries present~\cite{Pandey_2023, huang2015weyl, PhysRevLett.115.036806, PhysRevLett.115.036807}, or be fully gapped into topological insulators, depending on the strength of the SOC~\cite{yamakage2016line, PhysRevB.93.205132, schoop2016dirac, PhysRevB.93.201104, PhysRevB.93.201114, vanderbilt2018berry}. 
Since NLSMs exhibit rich physical properties and have potential applications in several fields of solid-state physics and materials science, it is natural to ask how robust the nodal line is against changes in SOC. Enhancing the robustness of nodal lines is an important challenge, especially in those materials where SOC plays a vital role. The potential methods have been proposed to enhance the robustness of nodal lines, including applying external electric or magnetic fields, substituting materials, applying local strain, and creating superlattices~\cite{yuan2013zeeman, PhysRevB.80.121308, PhysRevX.9.021028, PhysRevB.108.235166, PhysRevB.96.245101, pandey2025revisiting}. 
%%%%%%%%%%%%%%%%%%%%%%%%%%%%%%%%%%%%%%%%%%%%%%%%%%%%%%%%%%%%%%%
%%% What are the benefits Nodal line plane change
%%%%%%%%%%%%%%%%%%%%%%%%%%%%%%%%%%%%%%%%%%%%%%%%%%%%%%%%
Among these approaches, strain engineering is especially interesting in NLSMs, as it can effectively tune the band structure and potentially lead to exotic topological transport phenomena.
For example, applying external strain causes variations in the plane of the nodal line, which affects the directional dependence of the topological properties. In this context, the term external strain refers to a mechanical strain that physically alters the lattice constant, proving to be a powerful approach for tuning the physical properties of materials.
Furthermore, a recent study shows that uniaxial strain can induce a topological phase transition from a type-II to a type-III nodal ring, highlighting the tunability of topological phases via external strain~\cite{pandey2025revisiting}. In parallel, strain engineering has also been shown to tune the spin transport properties of quantum spin Hall systems, offering control over spin-dependent properties~\cite{zhang2017quantum, huang2017bending}.

%%% Which would be better single nodal v/s multinodal line

However, experimentally as well as theoretically, strain-dependent topological transport studies of NLSMs encounter obstacles that hinder a comprehensive investigation. 
One major obstacle lies in the Fermi surfaces of many real materials, which often exhibit multiple nodal rings rather than a single, isolated one.
These nodal rings can take various intricate forms such as linked nodal rings~\cite{PhysRevB.96.041102, PhysRevB.96.041103}, rectangular-shaped nodal-cages~\cite{bian2022visualizing}, and nodal-knot-shaped nodal rings~\cite{PhysRevB.96.201305}.
Such structures often become entangled in complicated manifolds, thereby adding complexity to the analysis. In addition to this structural complexity, the electronic dispersions in such systems are often quite complex due to the presence of many irrelevant trivial or nontrivial pockets with drumhead-like surface states at E$_{\rm{F}}$. 
This makes it particularly challenging to observe the spectroscopic and quantum transport signals directly attributable to the nodal lines. 
As a result of these intertwined challenges, most of the important physics and associated physical properties remain limited or have been difficult to access.
Consequently, materials that host a single nodal line are highly desirable and may be more effective than multi-nodal line systems in isolating transport signals associated with nodal line topology.

Compared to the widely studied type-II Weyl semimetals, stable materials hosting a type-II nodal-line phase with only a single nodal ring remain elusive, yet they offer a promising platform for enriching topological physics and enabling specific applications.
%Compared to the widely studied type-II Weyl semimetals, type-II NLSMs with only one nodal ring remain elusive, yet they offer a promising platform for enriching topological physics and enabling specific applications. 
For example, a type-II nodal line leads to anisotropic charge transport, which can be exploited in direction-dependent electronic devices~\cite{10.1063/1.5023320}. Additionally, this type of nodal line, with a strong tilt, exhibits novel properties such as direction-dependent chiral anomalies~\cite{PhysRevB.99.045143, extremelylargemagnetoresistancechiral} and highly anisotropic negative magnetoresistance~\cite{extremelylargemagnetoresistancechiral, PhysRevB.109.165155}, which can be leveraged in spintronic applications. 
Another promising feature of type-II NLSMs is the tilting of their nodal line bands, which leads to an asymmetric distribution of electron and hole carriers, gives rise to a large Nernst response, which enhances thermoelectric properties of the material~\cite{PhysRevB.105.115142, PhysRevMaterials.8.075403}. 
Building on this idea, Laksono \textit{et al.}~\cite{Laksono_2025} recently investigated the thermoelectric performance of type-II NLSMs. In their work, they demonstrate that optimizing the curvature of energy bands by tuning the Fermi velocity can significantly improve the thermoelectric performance of NLSMs.
In another study~\cite{PhysRevB.105.115142}, the thermoelectric transport properties of spinless Mg$_3$Bi$_2$ as a type-II NLSM were investigated. Their findings suggested that type-II NLSMs could represent a new class of high-performance thermoelectric materials, even among metals and semimetals. Taken together with the wide range of potential applications discussed earlier, these findings offer a guiding framework for the design of next-generation quantum materials with enhanced thermoelectric functionality.
Recently, the Cu$_2$SnS$_3$ has been identified to be type-II nodal line with only one nodal ring by the \textit{first-principles} calculations~\cite{pandey2025realization}. 
However, how this material can be beneficial from an application point of view is still unexplored, as is its response under uniaxial, equi-biaxial, and equi-triaxial strain. Strain can lead to flat energy dispersion, which provides an ideal pathway for many interaction-induced nontrivial phenomena. Therefore, it is meaningful to comprehensively investigate the effect of various strains on the nodal line phase of Cu$_2$SnS$_3$.

%%% Summary and main focus of the work
In this work, based on \textit{first-principles} calculations, we examine the robustness of the nodal line phase in Cu$_2$SnS$_3$ with respect to variations in the strength of SOC. Furthermore, we have investigated the effects of uniaxial, equi-biaxial, and equi-triaxial strains on the nodal line phase of Cu$_2$SnS$_3$. Remarkably, it has only non-trivial topological bands near E$_{\rm{F}}$, which provides a good base system to understand those properties completely associated with one nodal line~\cite{PhysRevResearch.4.033067}. 
Moreover, we have studied the intriguing physics that arises when strain is applied to nodal line system. Various reports have demonstrated the potential applications of such systems, including efficient electrocatalysts~\cite{li2024high}, high-performance thermoelectric converters~\cite{PhysRevB.105.085406}, thermionic devices~\cite{chen2020thermionic}, and spintronic devices~\cite{yang2021nodal}. Building on these findings, we propose that the strain-induced emergence of various nodal line phases could be harnessed for a wide range of practical applications, as discussed in the previous paragraphs.
The paper is organized as follows: In Sec. II, we present calculation methodology. In Section III(A), we demonstrated the robustness of the nodal line phase against changes in the strength of SOC. Next, in Section III(B), we explored the effects of uniaxial, equi-biaxial, and equi-triaxial strains and discussed the behavior of the nodal line, along with potential applications for different types of nodal line phases. We present the conclusions of our current work at the end of this paper.

%%%% Section 2 Computational details

\section{\label{sec:level2}Computational details}
\begin{figure}
\includegraphics[width=0.95\linewidth, height=5.5cm]{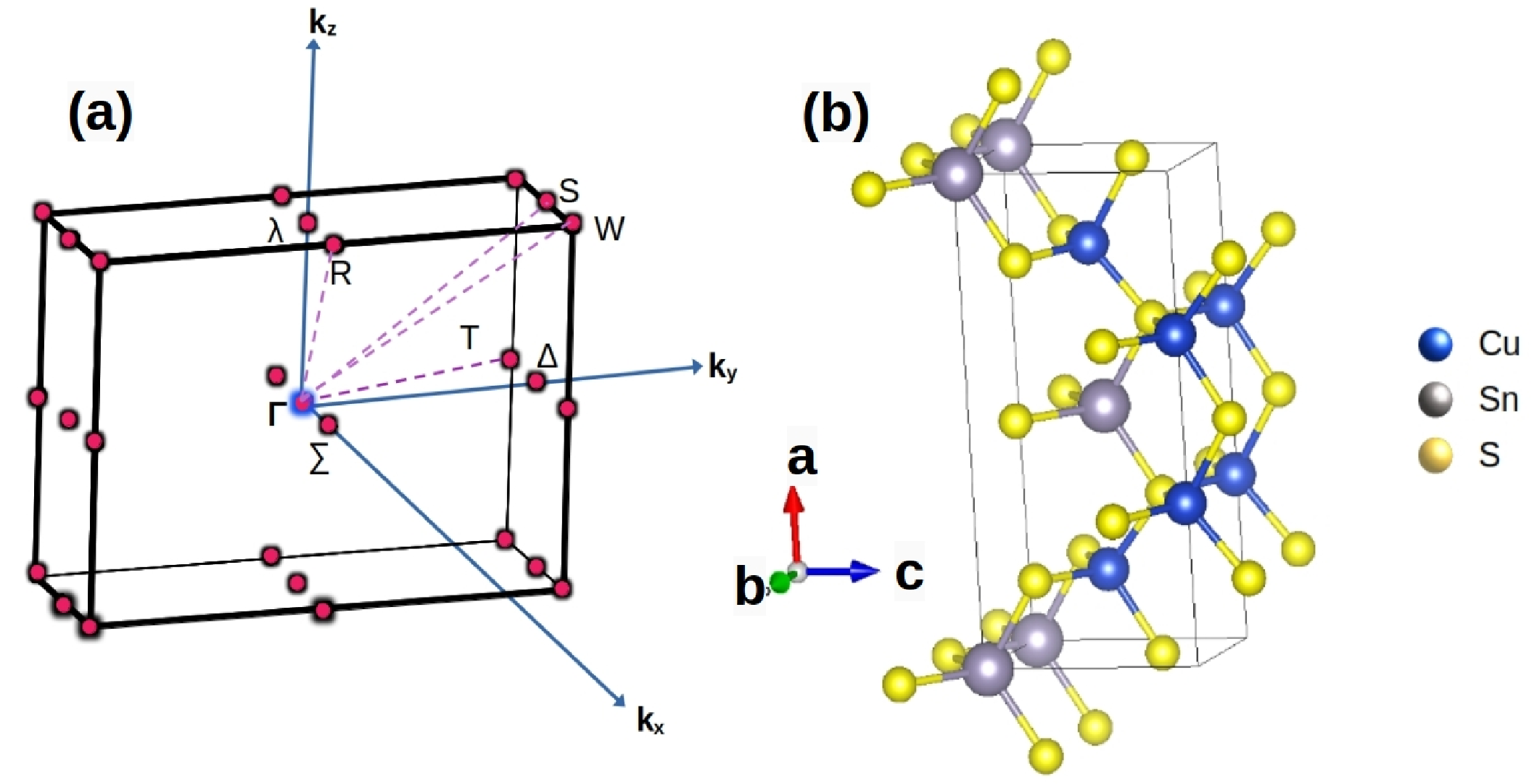}
	\caption{\label{fig:gimbal} \small{(Color online)	
(a) The Brillouin zone of a primitive unit cell of Cu$_2$SnS$_3$ with high symmetry points and paths indicated. (b) Crystal structure of Cu$_2$SnS$_3$.}}
\end{figure}

Within the formalism of the density functional theory (DFT), all the calculations are performed using \texttt{WIEN2k} package~\cite{blaha2020wien2k}. The electronic band structure of Cu$_2$SnS$_3$ is calculated using the augmented plane wave plus local orbitals (APW+lo) method. Generalized gradient approximation (GGA)~\cite{PhysRevLett.77.3865} within the Perdew-Burke-Ernzerhof is adopted as an exchange-correlation (XC) functional in all the calculations. This XC functional is used to optimize the geometry and lattice constant. The optimised lattice parameters are $a$=11.61 \AA, $b$=3.92 \AA, and $c$=5.43 \AA. The relaxed wyckoff positions of Cu, Sn, S1, and S2 are (0.8297, 0.5, 0.49053), (0.0, 0.0, 0.018978), (0.82306, 0.0, 0.26436), and (0.5, 0.0, 0.22018), respectively. In order to calculate the ground state energy of Cu$_2$SnS$_3$, 12$\times$12$\times$12 $k$-mesh size with the energy convergence limit $10^{-7}$ Ry/cell is taken. The analysis of nodal lines with different values of strain is performed using the PY-Nodes code~\cite{PANDEY2023108570}. This code is based on the Nelder-Mead's function-minimization approach~\cite{10.1093/comjnl/7.4.308, PANDEY2024109281}. 
The crystal structure and the corresponding Brillouin zone of the $Imm2$-phase of Cu$_2$SnS$_3$ are shown in Figs. \ref{fig:gimbal} (a) and \ref{fig:gimbal} (b), respectively.

\section{\label{sec:level2}Result and discussion}
\subsection{Robustness of the Nodal line phase against variations in SOC strength}
It is interesting to examine the extent to which the nodal line phase is robust with respect to the strength of SOC, as SOC typically leads to either a fully gapped phase or the evolution of the nodal line into several nodal points~\cite{yamakage2016line, PhysRevB.93.205132, PhysRevB.93.201104, PhysRevB.93.201114}. Therefore, we have performed electronic structure calculations with respect to different strengths of SOC to examine the robustness of the nodal line phase. Since the SOC Hamiltonian is inversely proportional to the square of the speed of light, decreasing the speed of light ($c$) could enhance the strength of SOC~\cite{blaha2020wien2k}. 
Therefore, $c$ is treated as a parameter, and its value is reduced in order to increase the strength of SOC in the calculations. In this work, we reduced the value of $c$ from 1370.3 to 548.1 a.u., which results in an increase in SOC strength from $\sim$ 0.0 to 1.9. Corresponding results are mentioned in Table \ref{table_soc}. It is important to mention here that the strength of SOC is calculated by taking the difference between the total energy of the Cu$_2$SnS$_3$ system obtained before and after the inclusion of SOC.
From the table, it can be seen that at zero SOC strength, Cu$_2$SnS$_3$ exhibits a nodal line phase. 
Even with an increase in SOC strength from 0.0 up to 1.5 meV, the nodal line phase persists.
Interestingly, when the strength is increased to 1.9 meV, the nodal line phase evolves into a Weyl phase. 
This result indicates that the 1.9 meV strength of SOC serves as a critical threshold, beyond which the nodal line phase vanishes and the Weyl phase emerges, remaining stable up to an SOC strength of 82.56 meV, corresponding to value of $c$=137.03 a.u.
The present material contains three elements: copper (Cu), tin (Sn), and sulfur (S), among which tin (Sn) has the highest atomic number ($Z$), i.e., 50. Given $Z$=50, the SOC effect is expected to be significant and unavoidable. Therefore, to reduce the strength of SOC, Sn atoms may be replaced with a Group 14 (IVa) element with a lower $Z$. This is because it is typically observed that in materials composed of light elements, the effect of SOC is weak~\cite{kaltsoyannis2006spin}; hence, maintaining the robustness of the nodal line phase is a feasible strategy. After substituting Sn atoms with lighter-mass atoms, the nodal line phase is expected to remain intact even higher strength of SOC and can be harnessed for further applications.

%%%%%%%%%%%%%%%%%%%%%%%%%%%%%%%%%%%%%%%%%%%%%%%
\begin{table}
\caption{\label{table_soc}
\small{The topological phase in Cu$_2$SnS$_3$ obtained corresponding to different strength of SOC per formula unit used in the calculations.}} 
\begin{ruledtabular}
\begin{tabular}{lccccc}
\textrm{{\textbf{S. No. }}}&
\textrm{{\textbf{SOC-strength}}}&
\textrm{\textbf{Topological Phase}}\\
\textrm{\textbf{}}&
\textrm{\textbf{(meV)}}&
\textrm{{}}\\

\colrule
   1       & 0.00     &  Nodal line \\
   2       & 0.50     &  Nodal line  \\
   3       & 1.00     &  Nodal line  \\
   4       & 1.50     &  Nodal line  \\         
   5       & 1.90     &  Weyl node   \\
  
\end{tabular}
\end{ruledtabular}
\end{table}

%%%%%%%%%%%%%%%%%%%%%%%%%%%%%%%%%%%%%%%%%%%%%%%%%%%%%%%%%%%%%%

\subsection{\label{sec:level2}Effect of axial Strain}
Since, in our earlier study~\cite{pandey2025realization}, we investigated the nature of the nodal line present in Cu$_2$SnS$_3$ at 0\% strain and found it to be type-II. Such a nodal line can be exploited for various potential applications, including large magnetoresistance effects and enhanced thermoelectric performance.
In quantum materials, using external perturbations, one may achieve a diverse range of topological phases. For example, uniaxial strain have recently proven to be effective and powerful methods for tailoring both physical properties and device functionalities, as reported in various studies~\cite{nicholson2021uniaxial, lin2024uniaxial, PhysRevB.109.035167, PhysRevLett.134.066501}. 
In our earlier study, we also investigated the effect of uniaxial compressive strain (UCS) of up to 6\% along \textit{a} direction only and found a type-III nodal line at 5.5\% UCS. 
Such nodal line features are associated with flat bands; hence, they serve as promising platforms for improved thermoelectric performance in materials.
Motivated by these findings, it would be interesting to investigate how uniaxial, equi-biaxial, and equi-triaxial strains influence the nodal line phase of this material, as well as to explore its potential practical applications under strain.
It is also important to mention here that in various reports, the topological properties have been studied under the application of uniaxial, equi-biaxial and equi-triaxial strains of up to 8\%~\cite{PhysRevLett.124.016402, Wang_2021, PhysRevB.98.045131, PhysRevResearch.4.023114, teshome2019topological, mondal2019emergence, 10.1063/5.0030200}. This naturally raises the question of whether such levels of strain are experimentally feasible. In this line, several experimental studies have also been reported, showing that up to 8\% external strain can be applied to the material~\cite{nie2019approaching, 10.1126/science.aat8051, tian2012approaching, 10.1126/science.1228602, PhysRevMaterials.1.033608}, confirming the practical feasibility of the theoretical predictions at the reported strain values.
Additionally, it is important to note that up to 8\% tensile strain on the lattice is also experimentally achievable, and various studies have been conducted in this regard~\cite{PhysRevLett.105.215502, 10.1126/science.aax9753, KIM20095245}. Hence, these studies motivate us to explore the behavior of the electronic properties under tensile strain.
In this direction, the electronic structure calculations are performed on the Cu$_2$SnS$_3$ system under uniaxial, equi-biaxial, and equi-triaxial strain. The corresponding band structures under various strains in different magnitudes and directions are given in Supplementary Material (SM)~\cite{spl}.

\begin{figure*}
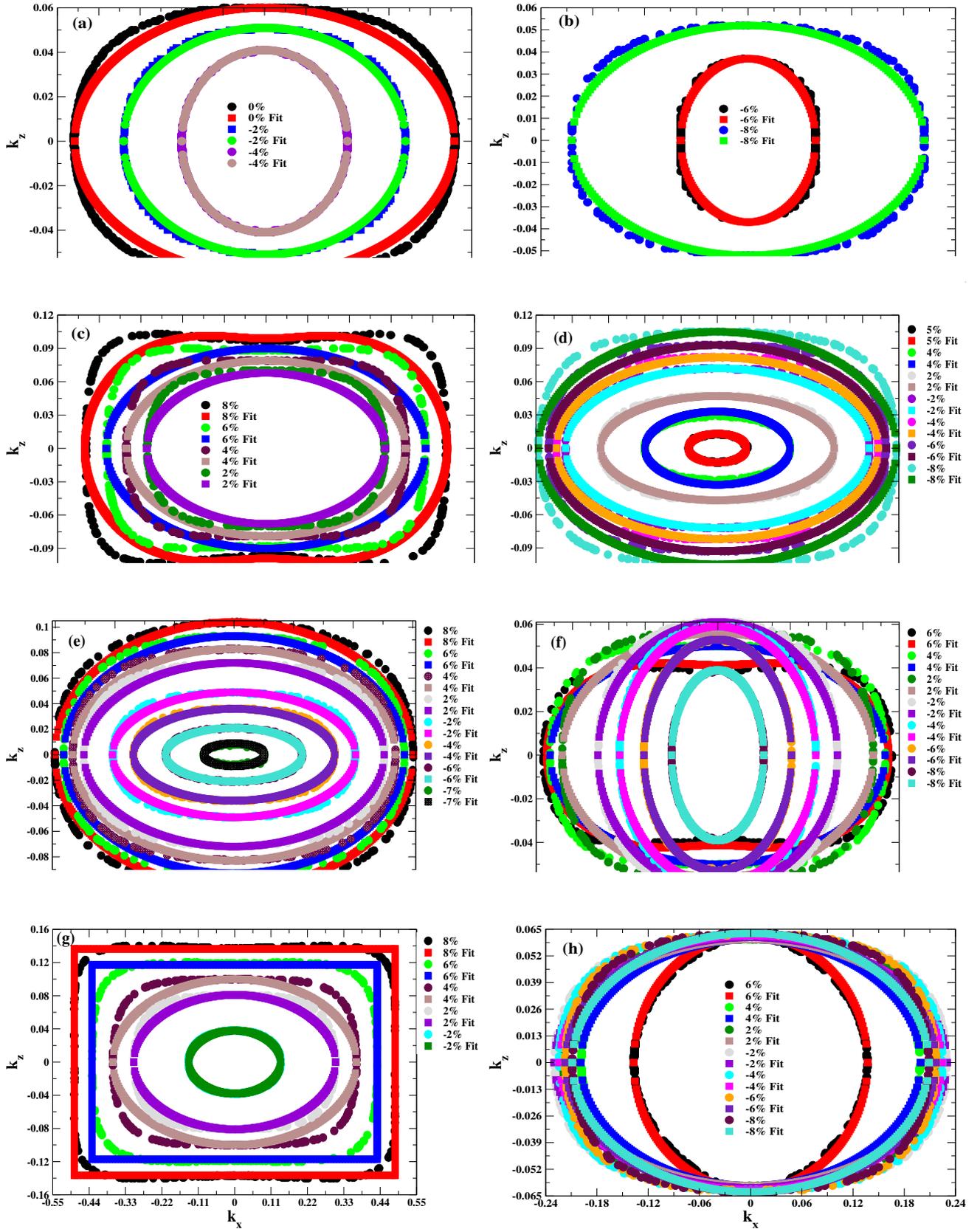

\includegraphics[width=0.48\linewidth, height=5.5cm]{0_4N_fit_a.eps}
\includegraphics[width=0.48\linewidth, height=5.5cm]{6n_10n_fit_a.eps}
\includegraphics[width=0.48\linewidth, height=5.5cm]{8P_2P_fit_a.eps} 
\includegraphics[width=0.48\linewidth, height=5.5cm]{5p_8n_fit_b.eps}
\includegraphics[width=0.48\linewidth, height=5.5cm]{8p_7n_fit_c.eps}
\includegraphics[width=0.48\linewidth, height=5.5cm]{6p_8n_ab_fit.eps}
\includegraphics[width=0.48\linewidth, height=5.5cm]{8p_2n_fit_ac.eps}
\includegraphics[width=0.48\linewidth, height=5.5cm]{6p_8n_fit_bc.eps}
\caption{\label{fig:8p_10n_a}\small{(Color online) 
Nodal line in Cu$_2$SnS$_3$ under various strains in different magnitudes and directions by means of the \textit{first-principles} approach. Fit describes different curves that are fit over the corresponding nodal line. The positive (negative) sign indicates a tensile (compressive) strain.
(a), (b), and (c) Along the \textit{a} direction, 2\% to 8\%, 0\% to -4\%, and -6\% to -10\%, respectively, a uniaxial strain is applied.
(d) 5\% to -8\% Uniaxial strain along \textit{b} direction. 
(e) 8\% to -7\% Uniaxial strain along \textit{c} direction. 
(f) 6\% to -8\% Equi-biaxial strain along \textit{a} and \textit{b} directions. 
(g) 8\% to -2\% Equi-biaxial strain along \textit{a} and \textit{c} directions. 
(h) 6\% to -8\% Equi-biaxial strain along \textit{b} and \textit{c} directions.}}
\end{figure*}

{\it Uniaxial Strain.|}
%%%%%%%%%%%%% Uniaxial Strain a
First, uniaxial strain is applied by varying the \textit{a} lattice parameter (from 0\% to $\pm$8\% along the \textit{a} direction, where the positive (negative) sign indicates tensile (compressive) strain) in steps of 2\%.
In this type of strain, the unit cell is strained in the \textit{a} direction only. For details about the band structures, see the SM~\cite{spl}.
In the case of uniaxial tensile strain (UTS) applied along the \textit{a} lattice parameter, for $0\% < \mathrm{UTS} \leq 8\%$, the top-most valence band (TMVB) near the $\Gamma$ high-symmetry point (along the X-$\Gamma$ path) shifts from below to above the E$_{\rm{F}}$, compared to the band structure of the unstrained lattice.
However, bottom-most conduction band (BMCB) along the $\Gamma$-S-T-X high-symmetric path shifted from above to below near E$_{\rm{F}}$.
While in the case of uniaxial compressive strain (UCS), for 0\%$<$UCS$<$5.5\%, the behavior of the band structure is almost similar as the band structure of unstrained lattice.
On the application of UCS, TMVB remains dispersive $<5.5$\%, which further evolves into flat band around the $\Gamma$ point for 5.5\%$\leq$UCS$<$5.6\%. 
With the further application of 8\% UCS, the situation becomes markedly different from the above discussed features of the band structure. 
At this strain value, the TMVB along the $\Gamma$-A high-symmetric path shifted from below to above the E$_{\rm{F}}$.
Therefore, the geometry and behavior of the nodal line are expected to vary with different strain values. To observe the variation in the behavior of the nodal line under strain, we have studied its response to both UCS and UTS.
In this context, we have applied UCS up to 8\% on the material and found very interesting and fascinating results. First, we observed that for UCS up to 4\%, the size of the nodal line decreases with increasing compressive strain along \textit{a} direction, as shown in Fig. \ref{fig:8p_10n_a} (a). Second, upon further application of UCS, we found that for 6\%$\leq$UCS$\leq$8\%, the size of the nodal line increases, but the number of nodal lines remains fixed. However, another interesting aspect is that the plane of the nodal line changes from $k_{{x}}$-$k_{{z}}$ ($k_y$=0) to $k_{{y}}$-$k_{{z}}$ ($k_x$=0) within this studied range of UCS, as evident from Fig. \ref{fig:8p_10n_a} (b). Thus, the strain-induced variation in the plane of the nodal line suggests that this material could be utilized to exploit the directional dependence of topological properties for example, in gated topological transistors, where current anisotropy is governed by the orientation of the nodal line~\cite{gilbert2021topological, 10.1002/adfm.201904784}.
Now, let us focus on UTS along \textit{a} direction. In this situation, we found that within the range of 2\%$\leq$UCS$\leq$8\%, the size of the nodal line increases with increasing tensile strain along \textit{a} direction. However, the number of nodal lines remains fixed (i.e. one), and the nodal line lies in the $k_{x}$-$k_{z}$ plane, as evident from Fig. \ref{fig:8p_10n_a} (c). 

%%%%%%%% Why is the Change in the Area of a Topological Nodal Line is so Important?
The above study demonstrates that strain can influence both the size and the very existence of the nodal line.
Hence, it is essential to understand why variations in the area of a topological nodal line (increasing or decreasing) are crucial from an application perspective. The change in the area of nodal line under strain directly affects the topological and transport properties of materials. 
For example, when strain is applied on the lattice, it changes the area of nodal line that (i.e., changes in \textbf{k}-space regions). 
This strain-induced modification leads to an unconventional Fermi surface geometry, which is expected to change the optical response of the material.
%%%%%%%% Why you have fitted the curve over theoretical result?
Along with this, we have also explored the geometry of the nodal line within the studied range of UCS and UTS, as the shape of nodal line (circular, elliptical, square, flower-shaped, etc.) is not just a mathematical construct but it plays a crucial role in determining topological properties of nodal line systems. This is especially important for many potential applications. 
For example, it is seen that the circular nodal line gives rise to a van Hove singularity near the Fermi level, with logarithmic divergence in the density of states (DOS)~\cite{PhysRevB.104.L041105, PhysRevB.111.195152, PhysRevB.108.235148}, which spawns effects such as superconductivity and charge and spin density waves, and leads to intriguing physics.
Furthermore, the elliptical nodal line leads to different effective masses and velocities along the axes, which affects the anisotropic optical conductivity~\cite{PhysRevB.96.155150}. Motivated by the presence of multiple nodal line shapes and their potential to host exotic quantum states that realize novel physics, various types of curves are fitted to the theoretical data. The equations corresponding to the fitted curves are available in the SM~\cite{spl}.

%%%%%%%%%%%%% Uniaxial Strain b
Next, we applied uniaxial strain to the \textit{b} lattice parameter to investigate the behavior of dispersion curves and the corresponding nodal line around E$_{\rm{F}}$. The dispersion curves are provided in SM~\cite{spl}. 
First, we applied UCS in the range 0\%$<$UCS$\leq$8\% to the material and observed that the size of the nodal line increases with increasing compressive strain along the \textit{b} direction, as evident from Fig. \ref{fig:8p_10n_a} (d).
However, upon applying UTS in the range of 0\%$<$UTS$\leq$5\% along the \textit{b} direction, the size of the nodal line decreases with increasing tensile strain. Moreover, the plane of the nodal line ($k_{y}$=0) remains invariant within the studied range of UCS and UTS along the \textit{b} direction.
Interestingly, when the UTS exceeds 5.5\%, the nodal line completely vanishes and does not reappear at higher strain values, indicating a topological phase transition from a nodal line phase to a gapped phase. At this strain value, the present material acts like a topological on-off switch, because low strain leads to a nodal line phase, which evolves into a trivial insulator at higher strain values. This behavior of the nodal line can also be leveraged in reconfigurable topological circuits~\cite{krishnamoorthy2023topological, zhang2018topological}.
%%%%%%%%%%%%% Uniaxial Strain c
Now, we focus on uniaxial strain applied to the \textit{c} lattice parameter. We found that the effect of UTS (UCS) on the dispersion curve is similar to that of UCS (UTS) along \textit{b} direction (see details in the SM~\cite{spl}). 
Therefore, the nature of the nodal line due to UTS (UCS) along the \textit{c} direction is expected to exhibit similar characteristics as UCS (UTS) along \textit{b} direction. 
To examine the nature and behavior of the nodal line under uniaxial strain, we have demonstrated the nodal line phase in the presence of UCS and UTS along the \textit{c} direction. 
We observed that for 0\%$<$UCS$\leq$7\% (0\%$<$UTS$\leq$8\%), the size of the nodal line decreases (increases) with increasing compressive (tensile) strain along the \textit{c} direction, as shown in Fig. \ref{fig:8p_10n_a} (e).
Notably, upon further application of UCS beyond 7\%, we found that the nodal line completely vanishes and does not reappear at higher levels of UCS.
Finally, from the above discussion, we remark that uniaxial strain in this material not only triggers the plane change but also induces various topological phases while keeping the number of nodal lines fixed. Thus, such a study highlights the importance of uniaxial strain in driving topological phase transitions through a single external parameter, which can be harnessed for various practical applications, including straintronic logic gates, optical switches, and topological quantum computing in the development of future electronic devices~\cite{gilbert2021topological, jin2023topological}.

\begin{figure}
\includegraphics[width=0.95\linewidth, height=6cm]{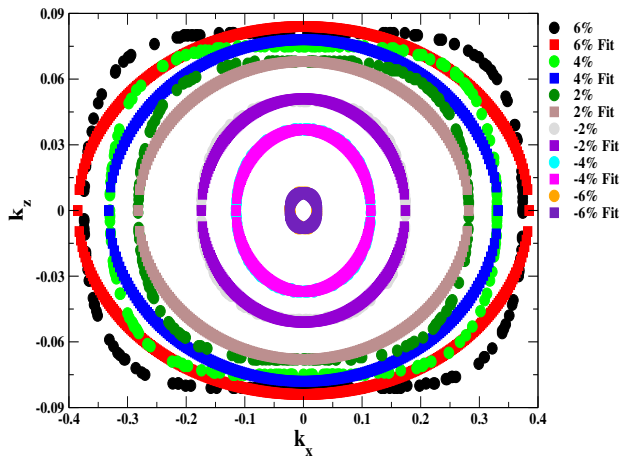}
\caption{\label{fig:6p_6n_fit_abc}\small{(Color online) 
Nodal line in Cu$_2$SnS$_3$ under 6\% to -8\% Equi-triaxial strain along \textit{a}, \textit{b} and \textit{c} directions by means of the \textit{first-principles} approach. Fit describes different curves that are fit over the corresponding nodal line.}}
\end{figure}

%%%%%%%%%%%%% Equi-biaxial Strain ab %%%%%%%%%%%%%%%%%%%%%%%%%%%%%%%%%%%%%%%%%%%
{\it Equi-biaxial Strain.|}
Now, we focus on the behavior of the nodal line under equi-biaxial strain. Firstly, we have applied the equi-biaxial strain along \textit{a} and \textit{b} direction only. 
Upon applying equi-biaxial compressive strain (EBCS) of up to 8\%, the TMVB near the $\Gamma$ high-symmetric point (along the X–$\Gamma$ path) shifted from above to below E$_{\rm{F}}$, compared to TMVB of the unstrained structure (see details in the SM~\cite{spl}). 
In addition to this, as the EBCS increases from 0\%, the flatness of TMVB slightly increases, while BMCB changes from smooth to cusped in the vicinity of the $\Gamma$ point. 
However, in the case of equi-biaxial tensile strain (EBTS) up to 6\%, along the S-T-X path, the BMCB shifts from above to below E$_{\rm{F}}$, while along the X-$\Gamma$ path, it remains above E$_{\rm{F}}$ (see details in the SM~\cite{spl}). Additionally, along the X-$\Gamma$path, the TMVB resides above E$_{\rm{F}}$. Therefore, the geometry and behavior of the nodal line are expected to differ under EBCS and EBTS. 
For this, we have determined the nodal line under various magnitudes of strain. Within this context, we have applied EBCS up to 8\% on the material and found size of the nodal line decreases with increasing compressive strain along \textit{a} and \textit{b} direction, as evident from the Fig. \ref{fig:8p_10n_a} (f). Moreover, the plane ($k_{{y}}$=0) and the number of nodal lines remain robust within the studied range of EBCS. In the case of EBTS, we found that within the range of 0\%$\leq$EBTS$\leq$6\%, the size of the nodal line increases with increasing tensile strain. Interestingly, with EBTS along \textit{a} and \textit{b} direction, the major axis of the nodal line increases, while the minor axis decreases monotonically. This is another distinct feature of this type of strain that has not been found in the case of uniaxial strain. Upon further application of 8\% EBTS, we found very interesting and fascinating results. At this strain value, one nodal line evolves into three nodal lines: one in the $k_{{x}}$-$k_{{z}}$ ($k_{{y}}$=0) plane and two smaller ones in the $k_{{y}}$-$k_{{z}}$ ($k_{{x}}$=0) plane (see details in the SM~\cite{spl}). 
These findings are entirely new, as uniaxial strain induces a change in the plane of the nodal line but does not lead to the emergence of a new nodal line, unlike the case of equi-biaxial strain in this material. 
The emergence of a new nodal line under this type of strain is expected to provide additional low-energy transitions, thereby modifying the joint DOS~\cite{10.1002/aelm.201900860, fumega2021increasing, schilberth2025generation}. 
Based on these findings, we propose that such features are expected to enable future developments in multifrequency light modulators, strain-tunable bandwidth photodetectors, and multi-channel topological transport devices~\cite{meng2023multi, thai2021mos2, ma2023multichannel}.

%%%%%%%%%%%%% Equi-biaxial Strain ac
Let us now focus on the behavior of the nodal line in the presence of EBCS and EBTS along the \textit{a} and \textit{c} direction only. 
To examine the effect of strain on the nodal line, we have performed calculations under various magnitudes of strain (see Fig. \ref{fig:8p_10n_a}(g) for the behavior of the nodal line under EBCS and EBTS). 
With the application of EBCS up to 3.5\%, the nodal line phase remains present. However, beyond 3.5\%, the nodal line completely vanishes and does not reappear at higher values of EBCS. Moreover, for 0\%$\leq$EBCS$\leq$3.5\%, the plane ($k_{{y}}$=0) and the number of nodal lines remain unchanged. 
Interestingly, in the case of EBTS, for 6\%$\leq$EBTS$\leq$8\% along \textit{a} and \textit{c} direction, one nodal line evolves into seven: three in the $k_{{x}}$-$k_{{z}}$ ($k_{{y}}$=0) plane (one large and two smaller), two smaller ones in the $k_{{y}}$-$k_{{z}}$ ($k_{{x}}$=0) plane, and two smaller ones in the $k_x$-$k_{y}$-$k_{{z}}$ plane (see Fig. 8 in the SM~\cite{spl}). 
Notably, within the seven nodal lines, we have found two sets of three mutually orthogonal, intersecting nodal loops (topological gimbals). It is important to mention here that a gimbal refers to a pivoted support that allows an object to rotate about an axis. This is another interesting and fascinating finding that has not been found in any of the previously discussed cases. 
Gimbal has various applications across multiple areas due to their capability to maintain orientation, such as in aircraft instrumentation (in gyroscopes)~\cite{sofka2007new, barbour1992inertial}, navigation and marine systems (in shipboard compasses)~\cite{grewal2007global, kayton2002navigation}, and photography and videography (for cinematic shots)~\cite{gavsparovic2017gimbal, rajesh2015camera}. In the context of topological phases, gimbal nodal rings give rise to a nontrivial Berry phase along different \textbf{k}-directions, which can be leveraged similarly to mechanical gyroscopes, such as quantum gyroscopes~\cite{PhysRevLett.93.137403, delgado2002quantum}. 
In addition to this, such nodal ring features can also be used in phase interference devices such as quantum interferometers~\cite{PhysRevA.74.051801, didomenico2004quantum}.
These observations strongly suggest that the EBCS alters the band topology, leading to the evolution of new topological nodal lines without requiring doping, chemical substitution, or any external perturbation.

%%%%%%%%%%%%% Equi-biaxial Strain bc
Along with this, we have also analyzed the nature and behavior of the nodal line in the presence of EBCS and EBTS along the \textit{b} and \textit{c} directions. It is found that the behavior of the dispersion curves due to EBCS is almost similar to that of the dispersion curves due to EBCS along the \textit{a} and \textit{b} directions. Meanwhile, the behavior of the dispersion curves due to EBTS is almost similar to that of the dispersion curves due to UTS along the \textit{c} direction. Therefore, the entire geometry of the nodal line is expected to change in the above-discussed cases in the presence of EBCS and EBTS. To examine this, the behavior of the nodal line at different strain values is studied. 
First, we applied EBTS up to 6\% on the material and observed that the size of the nodal line decreases with increasing tensile strain along the \textit{b} and \textit{c} directions, as evident from Fig. \ref{fig:8p_10n_a} (h). Moreover, similar to the results obtained from uniaxial strain along the \textit{a} direction, the plane of the nodal line ($k_{{y}}$=0) remains unchanged up to 6\%. Upon further application of EBTS beyond 6\%, the plane of the nodal line evolves from $k_{{y}}$=0 to $k_{{x}}$=0, while the number of nodal lines remains fixed at one. We have also found that within the range of 0\%$\leq$EBTS$\leq$6\%, the major axis of the nodal line decreases, while the minor axis increases. In the case of EBCS, we found that within the range of 0\%$\leq$EBTS$\leq$8\%, the size of the nodal line remains almost constant with increasing compressive strain compared to strains applied in other directions. 
Thus, this investigation demonstrates the power of equi-biaxial strain in driving the evolution of new topological nodal lines, opening the possibility of strain-tunable topological properties in the material.

%%%%%%%%%%%%% Equi-triaxial Strain abc %%%%%%%%%%%%%%%%%%%%%%%%%%%%%%%%%%%%%%%%%%%%%%%%%%%%%%%%%
{\it Equi-triaxial Strain.|}
Now, we turn on equi-triaxial strain along the \textit{a}, \textit{b}, and \textit{c} directions. Upon applying equi-triaxial tensile strain (ETTS) of up to 8\%, we find that the behavior of the dispersion curves under ETTS is almost identical to that under UTS along the \textit{a} direction. Meanwhile, the behavior of the dispersion curves under equi-triaxial compressive strain (ETCS) closely resembles that under EBCS along the \textit{a} and \textit{c} directions. Therefore, we expect that the geometry of the nodal line under equi-triaxial strain should exhibit features of both the nodal line under UTS and EBCS.
Within this context, we have applied ETTS up to 8\% on the material and found another exciting result. First, we observed that for ETTS up to 6\%, the size of the nodal line increases with increasing tensile strain along the \textit{a}, \textit{b}, and \textit{c} directions, as shown in Fig. \ref{fig:6p_6n_fit_abc}. While the nodal line expands with tensile strain, the number of nodal lines remains fixed and lies in the $k_x$-$k_z$ ($k_y$=0) plane. However, upon further application of ETTS, we found that for 6\%$\leq$ETTS$\leq$8\%, both the size and the number of nodal lines increase, rising from one to five (see Fig. 11 in the SM~\cite{spl}). The evolution from one nodal line to five new nodal lines, in which three lie in the $k_x$-$k_z$ plane and two in the $k_y$-$k_z$ plane, demonstrates the impact of ETTS. 
Such types of nodal lines give rise to anisotropic Berry curvature hot spots, increasing direction-dependent Hall responses which can be leveraged in strain-tunable Hall sensors and magnetoresistors~\cite{qi2023recent, qin2021strain, cenker2023strain, PhysRevMaterials.9.034412}. 
The evolution from a single nodal loop to multiple nodal loops under strain enables programmable electronic states that can be dynamically controlled. This tunability facilitates the development of straintronic devices with potential applications in both digital and analog systems~\cite{10.1063/5.0062993}.
In the case of ETCS, we found that within the range of 0\%$\leq$EBTS$\leq$6\%, the size of the nodal line decreases with increasing compressive strain along the \textit{a}, \textit{b}, and \textit{c} directions, as shown in Fig. \ref{fig:6p_6n_fit_abc}. However, the number of nodal lines remains one, and it lies in the $k_{x}$–$k_{z}$ plane within the range of 0\%$\leq$ETCS$\leq$6\%, as evident from the figure. Upon further application of ETCS above 6\%, the nodal line completely vanishes and does not reappear at higher values of strain. For ETCS above 6\%, the evolution from the nodal line phase to the gapped phase indicates that equi-triaxial strain may lead to topological phase transitions. 

Finally, we remark that applying external strain to Cu$_2$SnS$_3$ allows for richer topological nodal line structures including single nodal loops, nodal links, and intersecting nodal loops (gimbal) which could enable various device applications. For example, the type-II and type-III nodal lines can enhance thermoelectric performance; nodal lines lying in different planes may be utilized in multi-frequency light modulators, and multi-channel topological transport devices. %Moreover, topological gimbal-like nodal loops have potential applications in quantum gyroscopes.

\section{Conclusions}
In this work, the behavior of the nodal line in Cu$_2$SnS$_3$ under various strains of different magnitudes and directions is studied using a \textit{first-principles} approach. We found that applying uniaxial, equi-biaxial, and equi-triaxial tensile or compressive strain can effectively tune the electronic properties of Cu$_2$SnS$_3$, inducing topological phase transitions from a single nodal ring to more complex nodal structures. 
Depending on the magnitude and type of strain, we observed a rich variety of nodal line phases, including: (i) a single nodal line, (ii) three nodal lines, (iii) intersecting nodal loops (topological gimbals), (iv) five nodal lines, and (v) seven nodal lines. 
The tunable dispersion curves and robust nodal line phases make Cu$_2$SnS$_3$ a good candidate for both fundamental research and potential applications in various fields of solid-state physics and materials science.
The present work not only reveals a rich variety of topological nodal line phases in Cu$_2$SnS$_3$ in the presence of strain, but also helps in gaining a deeper understanding of multinodal line systems, spurring further experimental and theoretical work in the future.

\bibliography{MS}
\bibliographystyle{apsrev4-2}
\end{document}